# Development of a Wheelchair Simulator for Children with Multiple Disabilities

Nancy Rodriguez*

LIRMM (Laboratoire de Informatique, Robotique et Microéléctronique de Montpellier)

**ABSTRACT**

Virtual reality allows to create situations which can be experimented under the control of the user, without risks, in a very flexible way. This allows to develop skills and to have confidence to work in real conditions with real equipment. VR is then widely used as a training and learning tool. More recently, VR has also showed its potential in rehabilitation and therapy fields because it provides users with the ability of repeat their actions several times and to progress at their own pace. In this communication, we present our work in the development of a wheelchair simulator designed to allow children with multiple disabilities to familiarize themselves with the wheelchair.

**Index Terms**: I.3.1 [Computer Graphics]: Hardware Architecture — Input devices; I.3.7 [Computer Graphics]: Three-Dimensional Graphics and Realism – Virtual reality; H.5.1 [Information Interfaces And Presentation]: Multimedia Information Systems — Artificial, augmented, and virtual realities H.5.2 [Information Interfaces And Presentation]: User Interfaces — Input devices and strategies

## 1 INTRODUCTION

Training is an application area perfectly suited to virtual reality. VR allows the development of specific skills and to evolve independently VR simulation create a safe learning environment, user can test the consequences of his actions without risk in the real world. With a simulation, it is possible to expand the range of possible situations and to adapt them to different users, capabilities and knowledge[1][2]. In addition, the virtual environment allows to easy and quantitatively assessing the progress of the user and his weaknesses [3].

The objective of the current work we present in this article is the realization of an electric wheelchair simulator for children with multiple disabilities, aged between 13 and 16 years. This simulator will prepare children to wheelchair driving, taking into account motor and cognitive aspects. The first stage of this work concerns the development of a demonstrator to assess children's interest in the IT tool, which will then be integrated into their learning project.

---

* nancy.rodriguez@lirmm.fr

"Multiples disabilities" is a serious handicap associating motor impairment and severe mental retardation, resulting in a extreme restriction of autonomy and possibilities of perception, expression and relationship [4]. In fact, drive autonomously a wheelchair provides better mental and motor development by initiating new movements and by increasing socialization. But learning to drive a wheelchair in real conditions can be dangerous and tiresome sometimes even scary for patients, which can lead to slow this crucial stage in their development. The use of a virtual environment can help in this learning process and also allow them to use an electric wheelchair earlier, once children habituate and control the emotions provoked by driving.

In addition, the handling of the vehicle is not intuitive and requires practice, especially for patients with severe motor dysfunction preventing them from using conventional devices such as the joystick. A simulator may help find which devices are suitable for the individual and teach him how to use it without taking any risk.

This project is carried out in collaboration between the Institute for Children and Adolescents with multiple disabilities Coste Rousse located in Prades-le-Lez (Hérault, France) and the Laboratory of Informatics, Robotics and Microelectronics of Montpellier (LIRMM) in Montpellier. We work closely with the multidisciplinary team (educators, orthoptist, physiotherapist) of the center to make the simulator accessible from this conception.

## 2 RELATED WORK

Several wheelchair simulators are available; some designed to learn to drive or to evaluate the wheelchair and others to understand the difficulties of using a wheelchair and evaluate constructions accessibility. But there is not, to our knowledge, a simulator designed or adaptable for children with multiple disabilities. Marchal-Crespo presents in [5] a wheelchair prototype to facilitate the learning to drive, especially for the mentally handicapped. The system provides a chair with force feedback joystick and sensors to detect collisions. The prototype is able to avoid the walls and other obstacles to help movement.

In the work of Steyn[6], the simulator uses a immersive virtual reality interface displayed on 4 screens and a physical platform with a force feedback chair. ACCESSIM [7] uses a very realistic representation of urban areas to educate architects and decision makers on the difficulties faced by wheelchair users. The simulator VIEW [8]has a virtual reality interface and virtual chair is directed using a conventional joystick electric wheelchair (Figure 1). It is possible to visualize the virtual environment using a conventional screen or a helmet with track head movements.

Wheelsim[9] is a simulator constructed on the basis of a conventional racing game. Several courses (levels) are available. The duration of each course is recorded and the time spent outside the planned route. Note that a very important work was carried out on the sound environment in order to amplify immersion.

All these systems have realistic sets with many details and animations. This type of scene is not suitable for children suffering from cognitive disorders. Indeed, the details used to increase the realism and immersion of the user will cause difficulties in perception and understanding of the environment.

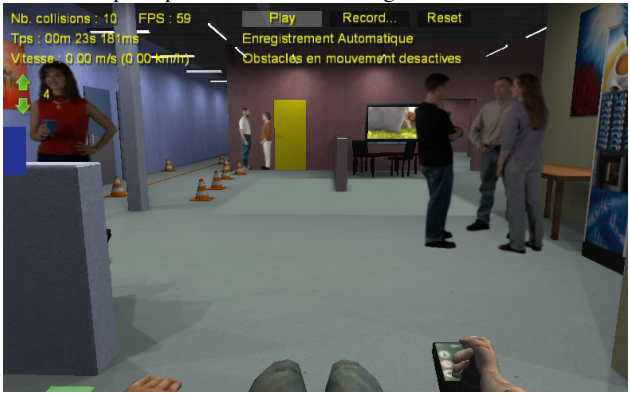

Figure 1: VIEW, a simulator for learning to drive

## 3 DESIGN

The simulator is developed under Unity3D, a game engine allowing to create and manage interactive virtual environments. The application provides many pre-programmed features and an ergonomic and intuitive interface. Unity offers several functionalities such as a physics engine, a tool for build animations and the possibility to integrate many input and output devices. Concerning the design and development of our prototype, all choices regarding look and sound, the different levels, gamification (positive reinforcement) and tests were carried out in consultation with the Coste Rousse team in a collaborative and iterative approach.

Central to the simulation is the virtual chair. It is the only object manipulated by the user and the goal is that its behaviour is fairly close to the behaviour of a real chair.

For rapid prototyping, we use a predefined model of a car wheels simulation. This simulation tool has four independent wheels, allowing simulating the behaviour of the wheels of various models of real wheelchairs. Of course, we also tested the behaviour of a wheelchair by using it, in order to better understand the feelings caused by its use. We then set the correspondence between the different movements of the wheels and the values of the joystick in order to better reproduce in the simulator the real behaviour of the wheelchair.

### 3.1 Visual aspects

Given the cognitive and perceptive troubles of children, it is essential to build a tool visually adapted. It is therefore necessary to identify the elements and codes that will allow conveying important information of the simulation. As we have stated above, the environments are usually very realistic with rich details for amplifying the immersion of the user. In our case, this approach does not make sense. Indeed, the accumulation of elements in a scene will tend to interfere with the user and prevent it from properly analyse the significant elements of the scene. We have therefore chosen to work on very refined environments by keeping only the necessary items. The colours have also been adapted to be sufficiently contrasted and then highlight the important objects and increase their visibility [10]. As a part of positive reinforcement, we decided to reward the player when he is on the right track with a discreet symbol (the Coste Rousse smiley) and colourful fireworks at the end of each level. Even if this effect is more intrusive, once the level is completed, the child's distraction is no longer a concern (see Figure 2).

At present, concerning sound feedback, there is a sound when the wheelchair starts to move and a sound (applause) associated with the fireworks at the end of the level. There is no yet haptic feedback in our system.

### 3.2 Interaction devices

A long-term goal of our project is the development of a software component to quickly integrate various interaction devices. This is a challenging task, since there is no standard for the input/output of of-the-shelf interaction devices, forcing to work on a case-by-case basis.

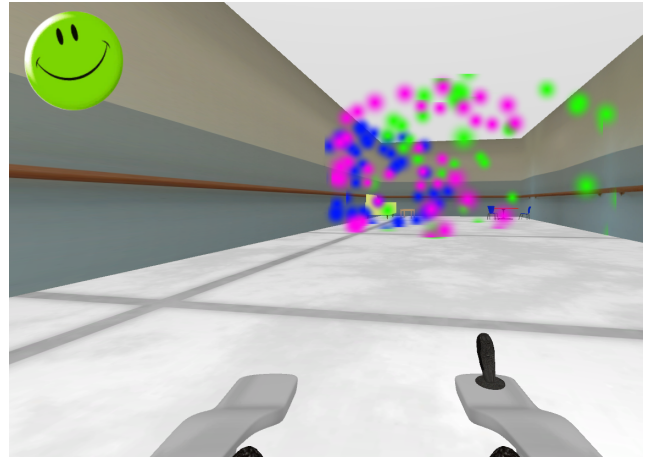

Figure 2: Visual feedback

We focused on the integration of a joystick, the default device on electric wheelchairs. After trying several joystick models, we decided to "build" our own joystick combining an unused joystick of a wheelchair (Figure 3) and a circuit board of a game joystick for translating data from analogic to digital format. This "hybrid" solution (a joystick from real wheelchair with a USB connexion) has enabled children to successfully control the virtual chair without being distracted by the interaction device.

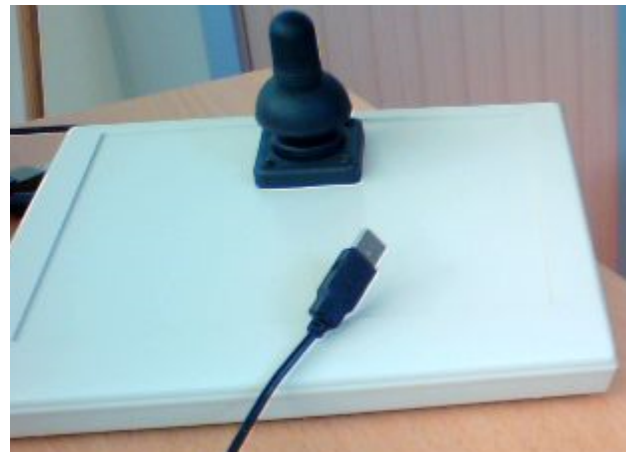

Figure 3: Joystick build for our simulator

## 4 Discussion

We have evaluated the prototype with four children of the Coste Rousse center, who know already drive a wheelchair, and have some computer skills. During various test sessions we observed a lot of curiosity about the simulator, and different ways of approaching it. Two children were focused on the screen and joystick while others tend to be more easily distracted. We did manage to get feedback on the recognition of objects in the virtual environment when the vehicle is stationary. However, recognition is more difficult in motion. The children managed to run the simulator for a while (2-3 minutes). But after a few minutes, we found difficulties in the use of devices and users wanted to stop. Indeed, the use of the joystick in a long time is difficult because of their mobility impairment and therefore it is necessary in the future to work with alternative devices.

In this project, the target users have extreme difficulties in communication; the level of verbal expression and even body language is very low. But studies and experiences in educational centers like Coste Rousse, show that children are able to express themselves and learn when you provide the right tools. Working with the team Coste Rousse and a small group of children, we managed to get ways to improve our work and offer a tool used by children with multiple disabilities. It is important to note that our system can only be tested with children in a advanced state of development. Indeed, application bugs of an early prototype usually provoke frustration and even rejection to use the simulator. For these reasons the phases of development and evaluation of our system are distant.

## 5 Conclusion

We have developed an electric wheelchair simulator suitable for children with multiple disabilities. We were able to experiment with a group of children and observed that it is an interest to add virtual reality in their learning project.

This ongoing work has allowed us to highlight the potential of virtual reality and challenges of such an approach in the field of health, to enable children with disabilities to grow in better conditions, develop new skills and express themselves. The keyword of our future work is "formalization". We now need to study various models of user centered design, interaction and evaluation in a context where users cannot communicate their needs and expectations. This will allow us, in collaboration with the multidisciplinary team in charge of children, to define an experiment enabling to evaluate the benefits of virtual reality for learning when users are experiencing multiple disabilities.